\documentclass[twocolumn,aps,superscriptaddress,nofootinbib,groupedaddress]{revtex4-1}

\usepackage{amsmath,amsthm,amsfonts}
\usepackage{graphicx}
\usepackage{physics}
\usepackage{hyperref}
\hypersetup{
	colorlinks   = true, %Colours links instead of ugly boxes
	urlcolor     = blue, %Colour for external hyperlinks
	linkcolor    = blue, %Colour of internal links
	citecolor    = red   %Colour of citations
}

\graphicspath{{figs/}}

\newcommand{\nc}{\newcommand}

\nc{\calR}{{\cal{R}}}
\nc{\calP}{{\cal{P}}}
\nc{\cN}{ {\cal{N}} }

\nc{\Mpt}{M_{_{\rm Pl}}^2}

\usepackage{tikz}
\usetikzlibrary{arrows,shapes}
\usetikzlibrary{trees}
\usetikzlibrary{matrix,arrows} 				% For commutative diagram
\usetikzlibrary{positioning}				% For "above of=" commands
\usetikzlibrary{calc,through}				% For coordinates
\usetikzlibrary{decorations.pathreplacing}  % For curly braces
\usepackage{pgffor}							% For repeating patterns

\usetikzlibrary{decorations.pathmorphing}	% For Feynman Diagrams
\usetikzlibrary{decorations.markings}
\tikzset{
    vector/.style={decorate, decoration={snake}, draw},
	provector/.style={decorate, decoration={snake,amplitude=2.5pt}, draw},
	antivector/.style={decorate, decoration={snake,amplitude=-2.5pt}, draw},
    fermion/.style={draw=black, postaction={decorate},
        decoration={markings,mark=at position .55 with {\arrow[draw=black]{>}}}},
    fermionbar/.style={draw=black, postaction={decorate},
        decoration={markings,mark=at position .55 with {\arrow[draw=black]{<}}}},
    fermionnoarrow/.style={draw=black},
    gluon/.style={decorate, draw=black,
        decoration={coil,amplitude=4pt, segment length=5pt}},
    scalar/.style={dashed,draw=black, postaction={decorate},
        decoration={markings,mark=at position .55 with {\arrow[draw=black]{>}}}},
    scalarbar/.style={dashed,draw=black, postaction={decorate},
        decoration={markings,mark=at position .55 with {\arrow[draw=black]{<}}}},
    scalarnoarrow/.style={dashed,draw=black},
    electron/.style={draw=black, postaction={decorate},
        decoration={markings,mark=at position .55 with {\arrow[draw=black]{>}}}},
	bigvector/.style={decorate, decoration={snake,amplitude=4pt}, draw},
}

\tikzstyle{block} = [draw, rectangle,
    minimum height=3em, minimum width=6em]

\begin{document}
	
\title{ PBHs and GWs from $\mathbb{T}^2$-inflation and NANOGrav 15-year data }

\author{Seyed Ali Hosseini Mansoori$^{1}$}
\email{shosseini@shahroodut.ac.ir}

\author{Fereshteh Felegary$^{1}$}
\email{fereshteh.felegary@gmail.com}

\author{Alireza Talebian$^{2}$}
\email{talebian@ipm.ir}

\author{Mohammad Sami$^{3,4,5}$}
\email{ sami\_ccsp@sgtuniversity.org }

%\author{Morteza Rafiee$^{1}$}
%\email{m.rafiee@shahroodut.ac.ir}

%\author{Alireza Tavanfar$^{6,7}$}
%\email{alireza.tavanfar@research.fchampalimaud.org}

\affiliation{$^{1}$Faculty of Physics, Shahrood University of Technology, P.O. Box 3619995161 Shahrood, Iran}
\affiliation{$^{2}$ School of Astronomy, Institute for Research in Fundamental Sciences (IPM), Tehran, Iran, P.O. Box 19395-5531}
\affiliation{$^{3}$Centre for Cosmology and Science Popularization (CCSP), SGT University,
Gurugram, Delhi- NCR, Haryana- 122505, India}
\affiliation{$^{4}$Eurasian International Centre for Theoretical Physics, Astana, Kazakhstan}
\affiliation{$^{5}$Chinese Academy of Sciences,
52 Sanlihe Rd, Xicheng District, Beijing}
%\affiliation{$^{6}$Champalimaud Research, Champalimaud Center for the Unknown, 1400-038 Lisboa, Portugal}
%\affiliation{$^{7}$Institute of Neuroscience, University of Oregon, Eugene, OR 97403, USA}
	
\begin{abstract}

In this paper, we propose a novel mechanism in $\mathbb{T}^2$-inflation to enhance the power spectrum large enough to seed  primordial black holes (PBHs) formation. To accomplish this, we consider the coupling function between the inflaton field and  $\mathbb{T}^2= T_{\mu \nu}T^{\mu \nu}$ term. PBHs formed within this scenario can contribute partially or entirely to dark matter (DM) abundance. Furthermore, the amplification in the scalar power spectrum will concurrently produce significant scalar-induced gravitational waves (SIGWs) as a second-order effect. In addition, the energy spectrum associated with SIGWs can be compatible with the recent NANOGrav 15-year stochastic gravitational wave detection and fall into the sensitivity range of other forthcoming GW observatories.
\end{abstract}

\maketitle %\pacs{98.80.Cq}

\section{Introduction}\label{sec0}
Recently, various collaborative efforts of Pulsar Timing Arrays (PTAs), such as NANOGrav \cite{NANOGrav:2023gor}, Parkers PTA \cite{Reardon:2023gzh}, European PTA \cite{Antoniadis:2023ott}, and the China PTA \cite{Xu:2023wog}, have collectively presented compelling evidence that firmly supports the existence of a stochastic gravitational wave background (SGWB) within the nHz frequency range. While the observed signal is predominantly attributed to standard astrophysical sources such as supermassive black hole binary mergers \cite{NANOGrav:2020spf, Middleton:2020asl}, it is worth considering the possibility that, in addition to the astrophysical background, the data might also has a cosmological origin. 

Among the potential cosmological interpretations of the SGWB are scalar-induced GWs (SIGWs) \cite{Vaskonen:2020lbd,DeLuca:2020agl,Kohri:2020qqd,Domenech:2020ers,Inomata:2020xad,Talebian:2022cwk,Frosina:2023nxu,Balaji:2023ehk}, first-order cosmological phase transitions \cite{Nakai:2020oit,Ratzinger:2020koh,NANOGrav:2021flc,Bringmann:2023opz,Ahmadvand:2023lpp}, as well as topological defects such as cosmic strings and domain walls \cite{Blasi:2020mfx,Ellis:2020ena,Ellis:2023tsl,Wang:2023len,Higaki:2016jjh}. For recent investigations on PTA results, please refer to Refs. \cite{Konoplya:2023fmh,Broadhurst:2023tus,Bian:2023dnv,Ge:2023rce,Anchordoqui:2023tln,Oikonomou:2023qfz,Zhu:2023faa,Bi:2023tib,Li:2023yaj,Han:2023olf,He:2023ado,Bhaumik:2023wmw,Ben-Dayan:2023lwd,Jiang:2023gfe,Bousder:2023ida} as well.

In recent years, there has been a growing research interest in SIGWs, which are generated as a second-order effect from the first-order scalar perturbations \cite{Ananda:2006af,Baumann:2007zm,Bugaev:2009zh,Assadullahi:2009nf,Alabidi:2012ex,Cai:2018dig,Pi:2020otn}. Importantly, if these scalar perturbations reach significant amplitudes on small scales, they can  give rise to a substantial population of PBHs \cite{Carr:2016drx,Carr:2021bzv,Ozsoy:2023ryl,Khlopov:2008qy,Belotsky:2014kca} and the SIGWs are simultaneously enhanced and can be sizable or even larger rather than the first-order GWs. 

On the other hand, we recently examine chaotic inflation within the context of the Energy-Momentum-Squared Gravity (EMSG) theory \cite{roshan2016energy}. The EMSG theory incorporates terms proportional to $\mathbb{T}^2\equiv T_{\mu \nu}T^{\mu \nu}$, where $T_{\mu \nu}$ is the energy-momentum tensor of the canonical scalar field Lagrangian \cite{HosseiniMansoori:2023zop}. In this respect, the EMSG will be a subset of the K-essence models \cite{armendariz2001essentials}. Essentially, to avoid ghost and gradient instabilities, specific restrictions must also be imposed on the model coupling parameter.

Despite recent observational bounds from Planck, WMAP, and BICEP/Keck during the 2018 observing season \cite{ade2021improved} ruling out chaotic inflation \cite{linde1983chaotic,linde1982new} with a potential of $\phi^{n}$ even for $n=2/3$ at approximately $95\%$ confidence level (CL), the presence of EMSG terms allows inflationary parameters, such as the spectral index $n_s$ and the tensor-to-scalar ratio $r$, to satisfy current observational constraints.

Over the past three decades, several mechanisms have been proposed to enhance the scalar power spectrum on small scales.
In the context of single-field inflation models, for instance, such an enhanced power spectrum can be achieved through imposing specific features on the inflaton potential such as a break in its first derivative \cite{Motohashi:2019rhu}, an inflection point \cite{Garcia-Bellido:2017mdw,Bhaumik:2019tvl}, and tiny bumps or dips in it \cite{Mishra:2019pzq}. 

 In this work, we introduce a novel mechanism aimed at significantly amplifying the power spectrum, resulting in the abundant production of PBHs. To be more precise, we focus on a model where a scalar field (inflaton) is coupled to the EMSG term. In fact, as the scalar-$\mathbb{T}^2$ coupling exhibits a rapid change during inflation, the curvature perturbations can be enhanced to become seeds of the primordial black holes
formed. It is interesting to note that in our model, these significant curvature perturbations not only result in the generation of primordial black holes, but also act as a source for second-order gravitational waves.
Furthermore, we try to show the possibility that the recent PTA data can be interpreted by the induced GW sourced by EMSG term during inflation.  Several other studies, for example \cite{Wang:2023ost,Ebadi:2023xhq,Liu:2023ymk,Cai:2023dls,Inomata:2023zup,Vagnozzi:2023lwo,Franciolini:2023pbf,Yi:2023mbm,Firouzjahi:2023lzg,Salvio:2023ynn,Choudhury:2023kam,Wang:2023ost,Datta:2023vbs,Bari:2023rcw}, have been carried out  on SIGWs as an explanation of the PTAs data.

This paper is organized as follows. Sec. \ref{sec1} begins by introducing  our setup within the EMSG framework. In Sec. \ref{sec2}, we discuss inflationary solutions in our scenario and show that the value of the spectral index $n_{s}$ and tensor-to-scalar ratio $r$ are compatible with the recent BICEP/Keck bound ~\cite{ade2021improved}. In the presence of the scalar-$\mathbb{T}^2$ coupling, we select some benchmark parameter sets and
discuss enhancements in the primordial curvature power
spectrum. Furthermore, we attempt to determine the fraction of PBH abundance in dark matter density at the present epoch. They will be further discussed in Sec. \ref{sec3}. In Sec. \ref{sec4} we investigate the possibility of detecting the energy spectrum of SIGWs from the recent NANOGrav signal and future GW experiments. Our conclusions are drawn in Sec. \ref{sec5}.

\section{Model}\label{sec1}
Allow us to consider the EMSG gravity action, which is given by \cite{roshan2016energy,HosseiniMansoori:2023zop}.
\begin{equation}\label{action}
S = \frac{1}{2}\int d^{4}x \sqrt{-g}\Big( M_{\text{\rm p}}^2 R  - M_{\text{\rm p}}^{-4} f(\phi) \mathbb{T}^{2}+2\mathcal{L}_{\rm m}\Big)
\end{equation}
where $M_{\rm p}$ is the reduced Planck mass, $R$ is the Ricci scalar associated with the spacetime metric $g_{\mu \nu}$, and $\mathcal{L}_{m}$ is the Lagrangian density corresponding to the matter source described by the energy-momentum tensor $T_{\mu \nu}$. Moreover, $\mathbb{T}$ is defined as $\mathbb{T}^2\equiv T_{\mu \nu}T^{\mu \nu}$ and $f(\phi)$ is a coupling function of inflaton $\phi$ \cite{HosseiniMansoori:2023zop}. It's important to note that in order to avoid ghost and gradient instabilities at the level of perturbations, it is crucial to ensure that $f(\phi)<0$ \cite{HosseiniMansoori:2023zop}. For convenience, we set $M_{\rm P}^2=1$ throughout this paper. 

Furthermore, in this model,  $T_{\mu \nu}$ is derived by varying the canonical scalar field Lagrangian  $\mathcal{L}_{\rm m}=X-V(\phi)$ where $X=-(\partial_{\mu} \phi \partial^{\mu} \phi)/2$ with respect to the metric, i.e.,
\begin{equation}\label{SM}
T_{\mu \nu}\equiv -\frac{2}{\sqrt{-g}} \frac{\delta(\sqrt{-g} \mathcal{L}_{m})}{\delta g^{\mu \nu}}=\partial_{\mu} \phi \partial_{\nu} \phi + g_{\mu \nu} ( X - V)
\end{equation}
Hence, one obtains
\begin{eqnarray}\label{T2}
\mathbb{T}^{2} = T_{\mu \nu}T^{\mu \nu} =  4 ( X^{2} - XV + {V}^{2}).
\end{eqnarray} 
By substituting the aforementioned result into the action \eqref{action}, the action can be rewritten as the K-essence \cite{armendariz2001essentials} model with a general function  $
P(X,\phi) =  \mathcal{L}_{\rm m}-f(\phi) \mathbb{T}^2/2$.
Here, we also consider the step-like coupling given by
\begin{equation}\label{coupling}
f(\phi)= \alpha+\mu_{1} \Big[\cosh\Big(\frac{\phi-\phi_{c}}{\mu_{2}}\Big)\Big]^{-2}
\end{equation}
where $\mu_{1}$, $\mu_{2}$, and $\phi_{c}$ are constants. Around the field value $\phi_{c}$, this coupling rapidly shifts to the asymptotic constant $f(\phi)=\alpha$. Hence, the value of $\phi_{c}$ can be determined by examining the evolution of $\phi$ in the $f(\phi)=\alpha$ model during the slow-roll inflation \cite{HosseiniMansoori:2023zop}. In the following section, we first review the cosmological predictions mentioned in Ref. \cite{HosseiniMansoori:2023zop}. Then, our focus will shift to estimating the quantity of $\phi_{c}$ at any given  number of e-folds ($N$).

\section{Chaotic slow-roll inflation with $f(\phi)=\alpha$}\label{sec2}
When we consider the slow-roll scheme, where $\dot{\phi} \ll V$ (or $X \ll V$) and $\ddot{\phi} \ll H \dot{\phi}$ (or $\dot{X} \ll H X$), we can derive the dynamical equation for the scale factor of the universe and the scalar field as follows \cite{HosseiniMansoori:2023zop}:
\begin{eqnarray}\label{Fridmannnew1}
3H^{2}&\simeq & V\left(1+2 \alpha V   \right)\\
\dot{\phi}V'\left( 1 + 4 \alpha V \right) & \simeq & -6 X H \left( 1 + 2\alpha V \right).\label{continuitynew1}
\end{eqnarray}
Clearly, Eq. \eqref{Fridmannnew1} reveals the presence of an upper bound on the potential, namely
\begin{equation}
V <\frac{1}{2 |\alpha|} 
\end{equation}
Note that in the above equation, we have considered the absolute value for $\alpha$. This choice is motivated by the findings of \cite{HosseiniMansoori:2023zop}, which demonstrate that in order to address the issues of ghost and gradient instabilities, the coupling constant $\alpha$ must be negative, specifically $\alpha<0$.

By differentiating both sides of Eq. (\ref{Fridmannnew1}) with respect to time and combining it with Eq. (\ref{Fridmannnew1}), we can derive the Hubble slow-roll parameter.
\begin{equation}\label{epsilonv1}
\varepsilon_{H} \simeq -\frac{1}{2 }\Big(\frac{V '}{V }\Big)\Big(\frac{\dot{\phi}}{H}\Big)\Big(\frac{1+4\alpha V }{1+2\alpha V }\Big).
\end{equation}
 By making use of Eq. \eqref{continuitynew1}, we can obtain
\begin{equation}\label{phidoth}
\frac{\dot{\phi}}{H} = -\Big(\frac{V '}{V }\Big)\Bigg[\frac{1+4 \alpha V }{\Big(1+2 \alpha
 V \Big)^2}\Bigg]
\end{equation}
As a result of the above relation, the slow roll parameter \eqref{epsilonv1} converts to
\begin{equation}\label{epsilonvnew}
\varepsilon_{H} = \frac{1}{2 }\bigg(\frac{V '}{V }\bigg)^{2}\Bigg[\frac{\Big(1+4\alpha  V \Big)^{2}}{\Big(1+2\alpha  V \Big)^{3}}\Bigg].
\end{equation}
Moreover, the Hubble slow-roll parameter $\eta_{H}$ is related to $\epsilon_{H}$ as
\begin{equation}
\begin{split}
\eta_{H}=\frac{\dot{\varepsilon_{H}}}{H \varepsilon_{H}}=\frac{\varepsilon_{H}'}{\varepsilon_{H}} \frac{\dot{\phi}}{H}
\end{split}
\label{etavnew}
\end{equation}
Notice that both slow roll parameters reduce to the standard form \cite{Li:2012vta} as $\alpha \to 0$. Additionally, these parameters must be much smaller than one, namely $\varepsilon_{H}, \hspace{0.25cm} \eta_{H}\ll 1$ during the inflation era which takes at least 50-60
number of e-folds to solve the flatness and the horizon problems. As a final remark, the inflation ends
when either of the slow-roll parameters tends to unity. 

Furthermore the sound speed, in the slow-roll limit, can be expressed as
 \begin{equation}\label{speed1}
 c_{s}^2=\frac{P_{,X}}{P_{,X}+2X P_{,XX}}\simeq 1+\frac{4\alpha  V}{3 } \Big(\frac{\dot{\phi}}{H}\Big)^2
 \end{equation} 
By taking advantage of Eq. \eqref{phidoth}, we can write down the above relation as a function of the potential and its derivatives.
The scalar and tensor power spectrum in the slow roll regime are also given by \cite{chen2007observational,Seery:2005wm}
\begin{equation}\label{slopower}
\mathcal{P}_{\mathcal{R}} \simeq \frac{1}{8 \pi^2} \frac{H^2}{\varepsilon_{H} c_{s}}|_{c_{s}k=aH}, \hspace{1cm}\mathcal{P}_{h}=\frac{2}{\pi^2} H^2|_{k=aH}
\end{equation}
Then, one can calculate the
spectral index $n_s$ and the tensor-to-scalar ratio $r$ as 
\begin{eqnarray}\label{nsanalatic}
n_{s}-1&\equiv & \frac{d \ln \mathcal{P}_{\mathcal{R}}}{d \ln k}\simeq -2 \varepsilon_{H}-\eta_{H}-s\\
r &\equiv & \frac{\mathcal{P}_{h}}{\mathcal{P}_{\mathcal{R}}}=16 \varepsilon_{H} c_{s}\label{rformula}
\end{eqnarray}
where $s\equiv \dot{c_s}/(H c_{s})$. 
 In the recent years, several observational constraints have been obtained on the $r$ and $n_{s}$ quantities from various data sources, including the \textit{Planck} 2018 data \cite{Planck:2018jri,BICEP2:2018kqh}, as well as BICEP/\textit{Keck} (BK15~\cite{aghanim2020planck} and BK18~\cite{ade2021improved}) data and BAO data. These limitations put serious restrictions on the free parameters of the model. 

Now, let us select a simple potential function like the chaotic potential with $V(\phi)=(A/M_{\rm p}^{n}) \phi^{n}$, where $A$ is a dimensionless coefficient and $n$ is a rational number. Note that  $A$ stands for the normalisation parameter given by the amplitude of the scalar power spectrum at the CMB pivot scale ($K_{\rm CMB}=0.05 \rm Mpc^{-1}$), i.e. $\mathcal{P}_{CMB} \sim 2.1 \times 10^{-9}$.

Using Eq. \eqref{phidoth}, the number of e-folding is also defined as 
\begin{equation}
N=-\int_{t_{e}}^{t} H dt=\int_{V_{e}}^{V} \frac{1}{2  \varepsilon_{H}} \frac{dV}{V} \Big[\frac{1+4\alpha V}{1+2\alpha V}\Big]
\end{equation}
 where the subscript ``$e$'' stands for the value of the quantities at the end of the inflation. Now by putting $V=A \phi^{n}$ into the above relation, we have
\begin{equation}\label{Nfunction}
N=\frac{1}{2 n}\Big(\frac{V}{A}\Big)^{\frac{2}{ n}}\Big[1+\frac{ 2\alpha V}{2+ n} \Big(1- {}_{2}F_{1}(1,1+\frac{2}{ n},2+\frac{2}{ n},-4 \alpha  V)\Big)  \Big] 
\end{equation}
  It should be noted that the potential $V$ in the above equation is calculated at the beginning of the inflation.
    Clealy, it is difficult to read the potential function $V$ as a function of $N$ in reverse. However, it can be accomplished if one chooses small values of the potential such that $|\alpha| V \leq \mathcal{O}(\sqrt{\varepsilon_{H}^{SC}})<1/2$ during inflation (the symbol $\rm SC$ represents the standard chaotic inflation.).  By assuming this and defining the expansion parameters as $\epsilon=\alpha A$, one can derive \cite{HosseiniMansoori:2023zop}
 \begin{eqnarray}
 \label{VAexpand}
\frac{V}{A}\simeq \Big(2n N\Big)^{\frac{ n}{2}}\Big[1-\Big(\frac{(2n)^{n+1} N^{n}}{(1+n)}\Big)\epsilon^2+\mathcal{O}(\epsilon^3)\Big] 
 \end{eqnarray}
While we have considered the above relation up to the second order, when $|\alpha| V$ approaches the bound $\sqrt{\varepsilon_{H}^{SC}} \sim \mathcal{O}(0.01)$ \cite{maldacena2003non}, it becomes necessary to consider higher orders, such as the fourth order, to achieve a strong agreement between analytical and numerical results.

By considering Eqs. \eqref{epsilonvnew}, \eqref{etavnew} and Eq. \eqref{VAexpand} together, we now derive the relation between the
 slow-roll parameters and $N$ as
 \begin{equation}\label{slowrollN}
\varepsilon_{H} \simeq  \frac{n}{4 N}\Big[1+2 (2n N)^{\frac{n}{2}} \epsilon-\frac{4 (2 n N)^{n} (1+2n)}{1+n} \epsilon^2
 +\mathcal{O}(\epsilon^3)\Big]
 \end{equation}
 \begin{equation}
\eta_{H}\simeq  \frac{1}{ N}\Big[1-n (2 n N)^{\frac{n}{2}} \epsilon+ \frac{2 n (2 n N)^{n}(3+5n)}{1+n} \epsilon^2
 +\mathcal{O}(\epsilon^3)\Big]
 \end{equation}
 The sound speed $c_{s}$ is also obtained to be
 \begin{equation}\label{csexpand}
 c_{s}\simeq 1+\frac{n}{3 N} (2 n N)^{\frac{n }{2}} \epsilon + \mathcal{O}(\epsilon^{2})
 \end{equation} 
 All above relations provide us with a formal solution for the spectral index \eqref{nsanalatic} as \cite{HosseiniMansoori:2023zop}
 \begin{eqnarray}\label{nsanaltic}
&&  n_{s}-1 \simeq  -\frac{1}{N}\Big[\frac{2+n}{2}-\frac{n(2nN)^{\frac{n}{2}} (n-2)}{6 N} \epsilon\\
\nonumber &+& \frac{n (2 n N)^{n}}{18(1+n)N^2} \Big(n(1+n)(n-2)+36 (2+3n)N^2\Big) \epsilon^2
 \\
\nonumber  &+&  \mathcal{O}(\epsilon^3)\Big]
 \end{eqnarray}
In comparison with Refs.  \cite{Li:2012vta,Unnikrishnan:2012zu},  the spectral index is modified by the orders of $\alpha A$. It is also trivial to derive $r$ as a function of $N$ when one combines Eqs. \eqref{rformula}, \eqref{slowrollN}, and \eqref{csexpand} together.

\begin{figure}
 \includegraphics[scale=0.6]{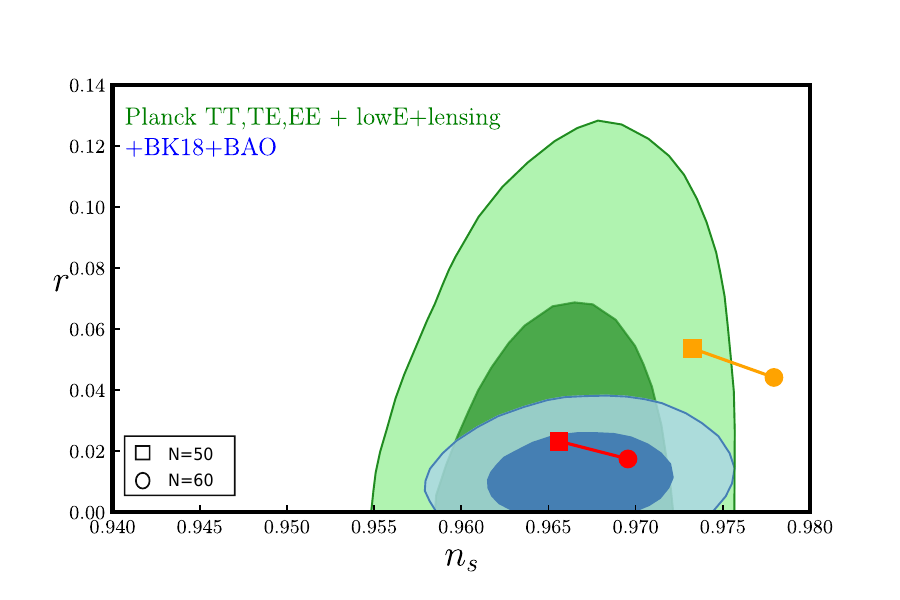}
 \caption{ Tensor-to-scalar ratio vs spectral index for EMSG model with the power law potential $V=A \phi^{2/3}$, compared to the data of Ref.~\cite{ade2021improved}. \label{Fignsr}}
\end{figure}

Fig. \ref{Fignsr} presents the tensor-to-scalar ratio as a function of the spectral index $n_{s}$ for $V=A \phi^{2/3}$, along with the observational constraints from the \textit{Planck} 2018 data, as well as BICEP/\textit{Keck} (BK15~\cite{aghanim2020planck} and BK18~\cite{ade2021improved}) data and BAO data. As seen, EMSG corrections result in improving the predicted values of $\{r,n_{s}\}$ in the standard chaotic inflation (orange shapes) such that they (red shapes) fall entirely within the region determined by the BK18 results~\cite{ade2021improved}.  

%%%%%%%%%%%%%%%%%%%%%%%%%%%%

 %%%%%%%%%%%%%%%%%%%%%%%

For instance,
the numerical values of $\{n_{s},r\}$ on CMB scales are $\{0.971802,0.0176263\}$ which are in close agreement with analytic result \eqref{nsanaltic} as one takes $\alpha A=-0.043$ and $N=60$.  
 In addition, by substituting $V=A \phi^{2/3}$ in Eq. \eqref{VAexpand}, the $\phi_{c}$ can be obtained as 
\begin{equation}\label{phic1}
\phi_{c}^{2/3} \simeq \Big(\frac{4}{3}N\Big)^{\frac{1}{3}}\Big[1-\frac{3}{5} \Big(\frac{4}{3}\Big)^{\frac{5}{3}}N^{\frac{2}{3}} (\alpha A)^{2}+\mathcal{O}\Big((\alpha A)^2\Big)\Big]
\end{equation}

\begin{figure}
\centering
 \includegraphics[scale=0.53]{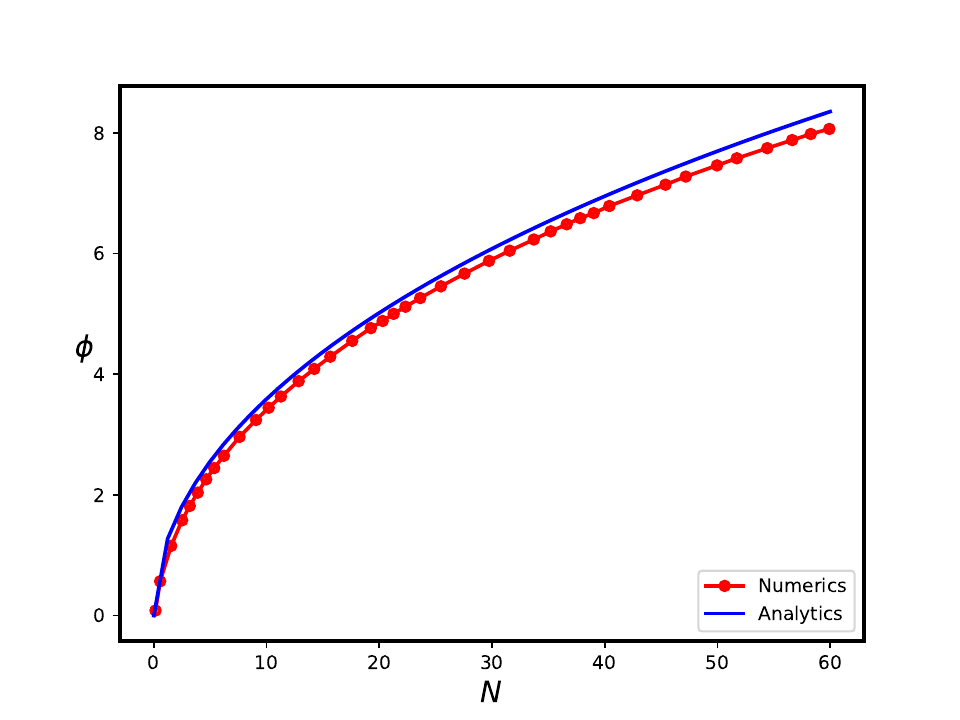}\hspace{0.1cm}
 \caption{The evolution of the scalar field during the inflationary phase. The red points indicate numerical results, whereas the blue solid curve is plotted by using \eqref{phic1} under the slow roll approximation. \label{Figphi}}
\end{figure}

In Fig. \ref{Figphi}, we compared the aforementioned result with numerical values represented by the red points. As shown, there is a satisfactory agreement between numerics and analytic results. In the presence of the coupling \eqref{coupling}, the background dynamics and perturbation spectra are subject to modifications. In the next section, similar to Refs. \cite{Kawai:2021edk,Zhang:2021rqs,Kawaguchi:2022nku}, we expect that this choice of the coupling can lead inflation into the ultra slow-roll (USR) stage and can thus significantly enhance the power spectrum $\mathcal{P}_{\mathcal{R}}$ of the primordial curvature perturbation on small scales, $k< k_{\rm CMB}$. As previously stated, the enhancement in the power spectrum of the scalar perturbations results in PBH formation with desirable masses and abundances. 

%%%%%%%%%%%%%%%%%%%%%%%
%%%%%%%%%%%%%%%%%%%%%%%%%%%%

%%%%%%%%%%%%%%%%%%%%%%%%%%%%%%%%%

\section{Power spectrum and PBH abundance
}\label{sec3}
\begin{figure}
\centering
 \includegraphics[scale=0.6]{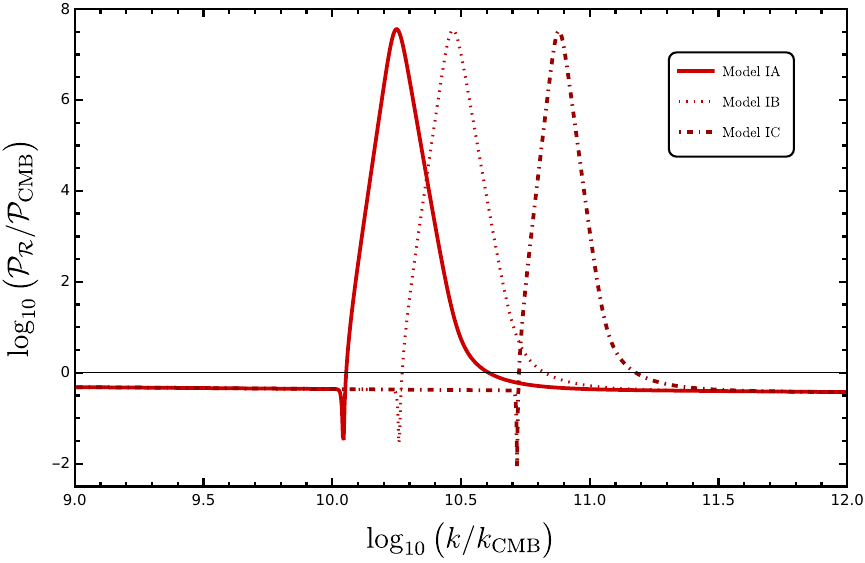}\\
 \includegraphics[scale=0.6]{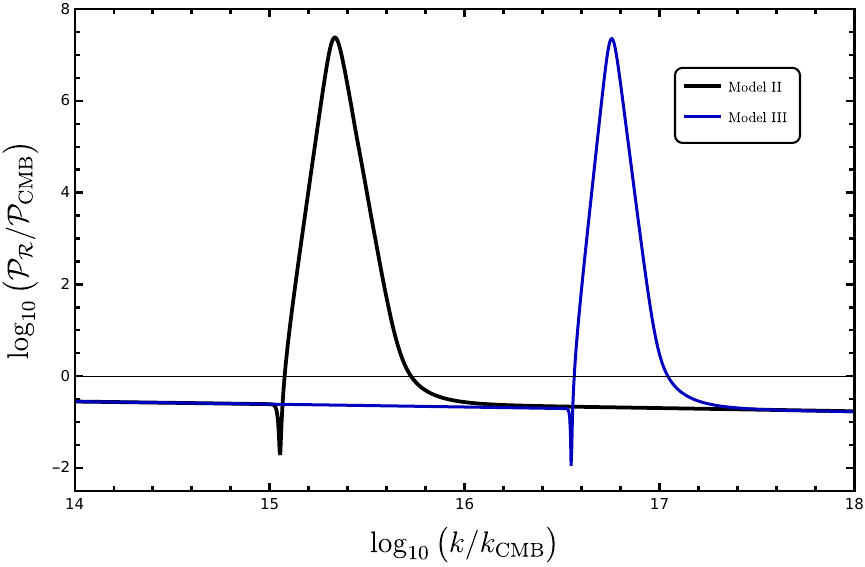}
 \caption{ Evolution $\calP_\calR$ for the models presented in Tab. \ref{tab1}.
		%The black dashed  curve presents the PBH bound~\citep{Garcia-Bellido:2016dkw}. 
%		Note that the green dashed  curve shows that the power can not be enhanced  when $f(\phi) = \alpha$.
 \label{Fig2}}
\end{figure}

In this section, we begin by calculating the power spectrum $\mathcal{P}_{\mathcal{R}}$ of the primordial scalar curvature perturbation using numerical methods. Afterward, we proceed to estimate the abundance of PBHs. Using the benchmark parameter sets listed in Tab. ~\ref{tab1}, we numerically generate the curvature perturbations power spectrum, as shown in Fig.~\ref{Fig2}. It's fascinating how adjusting the parameter space leads to a significant enhancement in the curvature power spectrum on smaller scales. As can be seen, the location of the peak is altered by the initial condition $\phi^{c}$\footnote{At the critical value $\phi_{c}$, the scalar field experiences a mild enough transition from the intermediate USR phase to the slow-roll (SR) phase where the curvature perturbations are enhanced. Therefore, as demonstrated in Ref. \cite{Firouzjahi:2023aum,Firouzjahi:2023ahg}, the smooth transition during the USR phase mitigates the concerns raised in \cite{Kristiano:2022maq,Riotto:2023gpm,Choudhury:2023jlt,Choudhury:2023rks} about the impact of loop corrections effects from small scale modes on CMB scale modes.}. In addition, the amplitude of ${\cal P}_{\cal R}(k)$ is controlled by $\mu_{1}$, while $\mu_{2}$ controls the width of the peak.

%%%%%%%%%%%%%%%%%%%%%%%
%%%%%%%%%%%%%%%%%%%%%%%%%%%%

 \begin{table}
 \begin{tabular}{l*{4}{c}r}
Model & \hspace{0.1cm} $\mu_{1}/10^{7}$ & \hspace{0.2cm} $\mu_{2}/ 10^{-3}$ & \hspace{0.2cm} $\phi_{c}$ & \hspace{0.2cm} $M_{\rm PBH}/M_{\odot}$\\
\hline
IA & \hspace{0.1cm} -1 &\hspace{0.2cm}  1.563 & \hspace{0.2cm}  6.598& \hspace{0.2cm}  $1.85 \times 10^{-1}$\\
IB & \hspace{0.1cm} -1 &\hspace{0.2cm}  1.610 & \hspace{0.2cm}  6.557& \hspace{0.2cm}  $7.0\times10^{-2}$\\
IC & \hspace{0.1cm} -2 &\hspace{0.2cm}  1.018 & \hspace{0.2cm}  6.47& \hspace{0.2cm}  $7.0 \times 10^{-3}$\\
II & \hspace{0.1cm}  -1.02 & \hspace{0.2cm}  3.512 & \hspace{0.2cm}  5.54& \hspace{0.2cm}  $2.4\times10^{-11}$ \\
III & \hspace{0.1cm}  -2.655 & \hspace{0.2cm}  2.1 & \hspace{0.2cm}  5.17 & \hspace{0.2cm}  $1.85\times10^{-14}$\\
%$t_{\rm scr}$            & 25.237 & 13.943 & 10.233 & 8.419 &  7.362 & 6.682  \\
%Benfica           & 6 & 2 & 1 & 3 &  7 & 8 &  7  \\
%FC Copenhagen     & 6 & 2 & 1 & 3 &  5 & 8 &  7  \\
\end{tabular}
\caption{Some model parameter sets. Note that the initial value for the scalar field in all models is $\phi_{0}=8.2 \text{M}_{\rm P}$ with $A=2.07\times 10^{-10}$. \label{tab1}}
 \end{table}
 
%Interestingly, for values of $\mu_{i}$ considered in Table~\ref{tab1}, the amplification in the power spectrum is large enough to seed PBH formation through the collapse of Hubble-size overdense regions after horizon re-entry of the fluctuation mode $k_{\rm PBH}$ ~\citep{Sasaki:2018dmp}. 

In addition, the enhanced power spectrum can lead to a significant contribution of PBHs to the DM density today. The fraction of PBHs against the total DM density at the present can be given by ~\citep{Sasaki:2018dmp}
\begin{equation}
f(M_{\rm PBH})= 2.7 \times 10^{8} \Big(\frac{0.2}{\gamma}
\sqrt{\frac{g_{*}}{10.75}} \frac{M_{\rm PBH}}{M_{\odot}}\Big)^{-1/2} \beta(M_{\rm PBH})
\end{equation} 
where the constant $\gamma$ measures how much fraction of mass transformed to be PBHs, $g^{*}$ is the relativistic degrees of freedom at formation, $M_{\odot}$ is the solar mass, and the $\beta$ is the mass fraction of PBHs at formation time.
The mass of the formed PBH as a function of a comoving scale $k$ is also given by \cite{Ballesteros:2017fsr}
\begin{equation}
\frac{M_{\rm PBH}(k)}{M_{\odot}}= 30 \Big(\frac{\gamma}{0.2}\Big)\big(\frac{g_{*}}{10.75}\Big)^{-1/6} \Big(\frac{k }{2.9 \times 10^{5} \text{Mpc}^{-1}}\Big)
\end{equation}
%where $k$ can be translated to the e-folds number $N$ via $k=k_{\text{CMB}}\exp(N_{\text{CMB}}-N)$.
%in which $H_{\rm end}$ and $H_{\rm PBH}$ are the Hubble rates at $N_{\rm end}$ and $N_{\rm PBH}$, respectively. 
According to the Press-Schechter formalism \citep{Press:1973iz}, the mass fraction $\beta$ for a given mass is defined as the probability that the Gaussian comoving curvature perturbation $\mathcal{R}$ (or the density contrast $\delta$) is larger than a certain threshold value $\mathcal{R}_{c}$ (or $\delta_{c}$) for PBH formation ~\citep{Press:1973iz,Lyth:2012yp,Byrnes:2012yx,Garcia-Bellido:2016dkw}. In this respect, by taking the Gaussian probability distribution function (PDF) for the curvature fluctuation spectrum, the fraction of collapsing regions at formation can be estimated as
\begin{equation}
\beta (k) \simeq \frac{1}{2} \rm Erfc \Big(\frac{\mathcal{R}_{c}}{\sqrt{2 \mathcal{P}_{\mathcal{R}}(k)} }\Big)
\end{equation}
 Recent numerical and theoretical investigations indicate that $\mathcal{R}_{c} \sim \mathcal{O}(1)$ ~\citep{musco2005computations,musco2009primordial,nakama2014identifying,harada2013threshold}.  In addition, the proper value of the threshold depends on the shape of the power spectrum of the curvature perturbation. In this paper, we take $\mathcal{R}_{c} \sim 1.75$ according to the amount of density threshold $\delta_{c} \sim 0.55$ quoted in \citep{musco2021threshold} by using the linear relation $\mathcal{R}_{c}=9/(2 \sqrt{2})\delta_{c}$ between curvature and density threshold \citep{drees2011running,young2014calculating,motohashi2017primordial}.

%%%%%%%%%%%%%%%%%%%%%%%
\begin{figure}
\centering
 \includegraphics[scale=0.48]{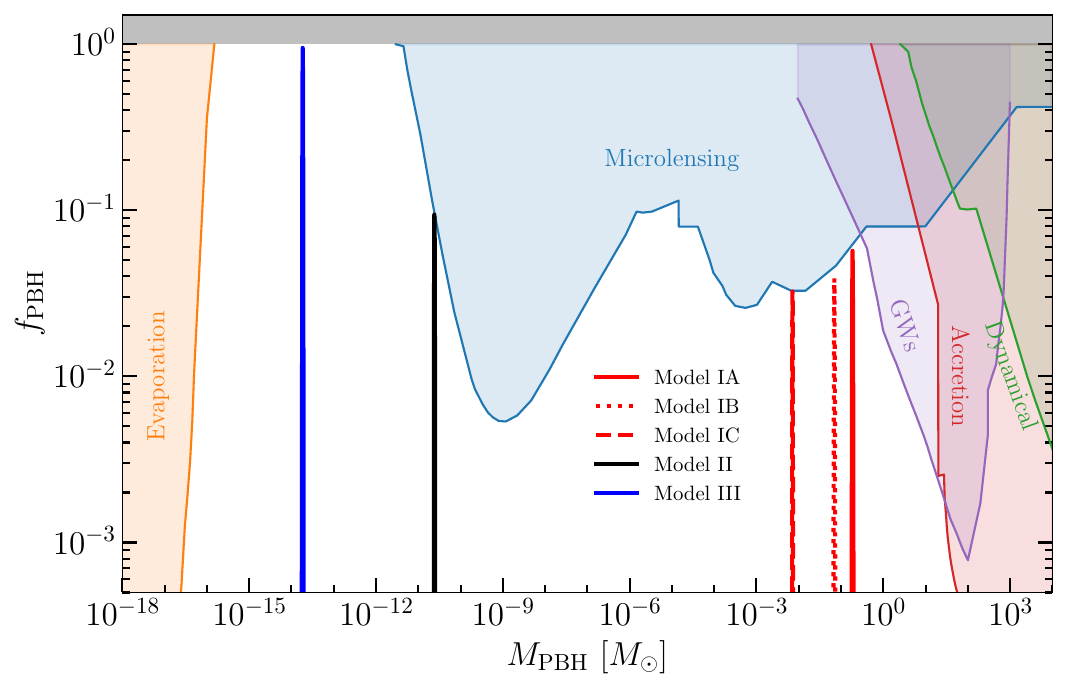}\hspace{0.1cm}
 \caption{ Fraction  $f_{\text{PBH}}$ as a function of the mass of the formed PBHs in the unit of solar mass for models in Table.~\ref{tab1}. The observational bounds are taken from Refs.~\citep{Green:2020jor,bradley_j_kavanagh_2019_3538999,Carr:2020gox}. . \label{Fig3}}
\end{figure}
%%%%%%%%%%%%%%%%%%%%%%%%%%%%

%The threshold for PBHs at the cosmological horizon crossing 
%has been widely computed in the literature  by making use of a linear extrapolation from the superhorizon regime. However, since the non-linear  relation between the density contrast and the curvature perturbation is neglected  it does not yield to the right amplitude of the perturbation at the cosmological horizon crossing \citep{Biagetti:2021eep,Musco:2018rwt}.  

%Taking into account this  non-linear effects, the mass fraction is defined as~\citep{Biagetti:2021eep}
%\begin{align}
%\label{beta-delta}
%\beta \equiv \int_{\delta_{l,c}}^{4/3} \dd \delta_l ~ \kappa
%\Bigg(
%\delta_l-\dfrac{3}{8}\delta_l^2
%-\delta_{c}
%\Bigg)^{\tilde{\gamma}}~f_{\delta_l}(\delta_l)
%\end{align}
%where the probability distribution $f_{\delta_l}(\delta_l)$ is given by \eqref{f_delta},  $\kappa=3.3$, and $\tilde{\gamma}=0.36$ for the collapse at the radiation-dominated epoch. Moreover, the field $\delta_{l}$ depends on the threshold density contrast through the following expression ( see App. \ref{PDF_delta} for more details).
%\begin{align}
%\delta_{l,c} = \dfrac{4}{3}\bigg(
%1-\sqrt{1-\dfrac{3}{2}\delta_{c}}
%\bigg)
%\end{align}
%in which one can take $\delta_{c} \simeq 0.59$ for a monochromatic curvature perturbation power spectrum~\citep{Musco:2018rwt, Musco:2020jjb}.

In Fig.~\ref{Fig3}, we have depicted $f_{\rm PBH}$ for the model parameters in Table~\ref{tab1}.  As illustrated, the formed PBHs can furnish  a large fraction of total DM abundance. In particular, for model III, we obtain $f_{\rm PBH} \simeq 1$ corresponding to $M_{\rm PBH} \sim 10^{-14}M_\odot$. 
%As mentioned before, a large enhancement in the amplitude of scalar perturbations results in the formation of primordial black holes. Moreover, such large curvature perturbations can act as a source for the second-order tensor perturbations, and thus generating SIGWs in the radiation domination (RD) era.
\section{Detectability of Induced Gravitational Waves by new results of NANOGrav}\label{sec4}
\begin{figure}
\centering
 \includegraphics[scale=0.245]{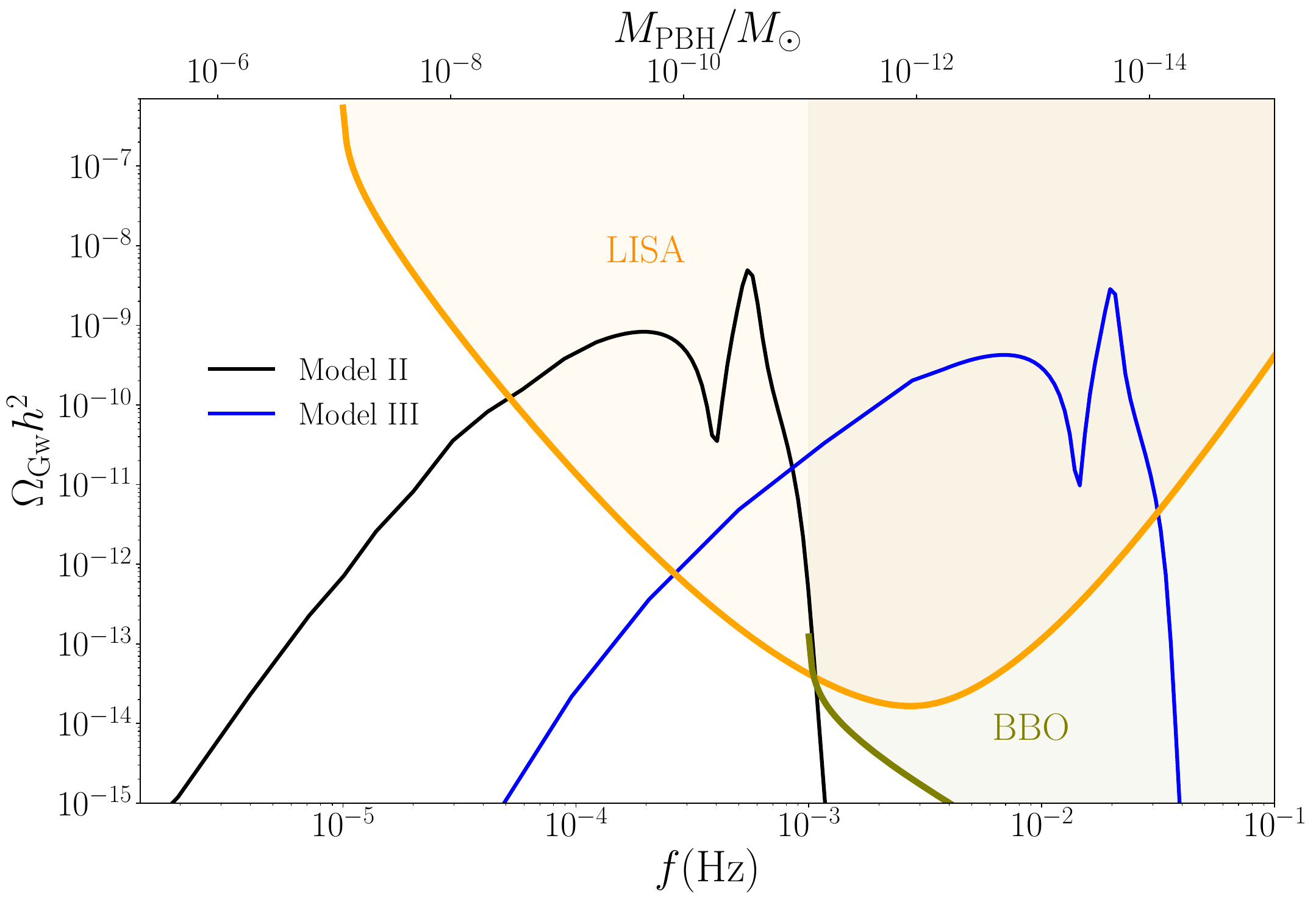}\hspace{0.1cm}
 \caption{ The energy density of the induced GWs for model II and III with respect to frequency. 
	The shadowed regions indicate the sensitivity curves of various GW detectors~\citep{Schmitz:2020syl,schmitz_kai_2020_3689582}.  
	\label{Fig4}}
\end{figure}

As mentioned, large curvature perturbations can act as a source for the second-order tensor perturbations, and thus generating SIGWs in the radiation domination (RD) era.
In this part, we therefore concentrate on the possibility of the enhanced scalar perturbation power spectrum as the source of the GWs within the new results of NANOGrav.
The energy density of the induced GW is given by \cite{Baumann:2007zm,Espinosa:2018eve}
\begin{eqnarray}
\nonumber  \Omega_{\rm GW}=\frac{\Omega_{r,0}}{36} \int_{0}^{1/\sqrt{3}} \rm d u \int_{1/\sqrt{3}}^{\infty} \rm d v \Big[\frac{(u^2-1/3)(v^2-1/3)}{v^2-u^2}\Big]^2&& \\
 \nonumber  \mathcal{P}_{\mathcal{R}}(\frac{k \sqrt{3}}{2}(v+u)) \mathcal{P}_{\mathcal{R}}(\frac{k \sqrt{3}}{2}(v-u))  \Big(I_{c_{1}}^2(v,u)+I_{c_{2}}^2(v,u)\Big) \,\ &&
\end{eqnarray}
where $\Omega_{r,0}\simeq 8.6 \times 10^{-5}$ is the radiation density at present and the functions $I_{c_{1},c_{2}}$ are defined in Appendix D of Ref. \cite{Espinosa:2018eve}. We also refer interested readers to Appendix D of Ref. \cite{Ozsoy:2020kat} in more detail. In Fig. \ref{Fig4}, we have plotted the quantity $\Omega_{GW} h^2$ in terms of frequency $f=k/2 \pi=1.55 \times 10^{-15} (k/1 \rm Mpc^{-1}) \rm Hz$ with $h^2=0.49$ together with the sensitivity of the various forthcoming
GW experiments \textit{e.g.} the Laser Interferometer Space Antenna (LISA) \cite{Bartolo:2016ami}, the Big Bang Observatory (BBO) \cite{Crowder:2005nr,Corbin:2005ny,Baker:2019pnp}.
Clearly, for model III, $\Omega_{\rm GW} h^2$ falls within the sensitivity of the BBO and LISA, while GWs for model II only peak well inside the range of detectability of LISA. 

%%%%%%%%%%%%%%%%%%%%%%%
%%%%%%%%%%%%%%%%%%%%%%%%%%%%

\begin{figure}
\centering
 \includegraphics[scale=0.6]{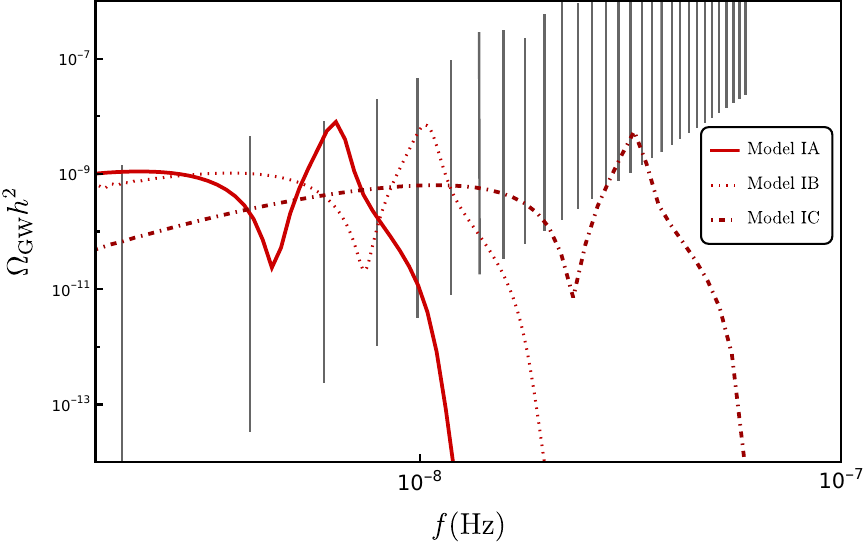}\hspace{0.1cm}
 \caption{ The energy density of the induced GWs for sets of the model I with respect to frequency. 
	The observation results of NANOGrav \cite{NANOGrav:2023gor} are presented with gray violins.  \label{FigNONGrav}}
\end{figure}

In Fig. \ref{FigNONGrav}, we have depicted the spectrum of SIGWs for Model I sets and compared them to the NANOGrave results.
 As seen, the energy density for all model I sets follow the NANOGrav 15-year results on the stochastic gravitational wave background. However, fully explaining the PTA signal with SIGW proves to be a challenging task due to the PBH bound in our model.

Taken together, these results suggest the PBHs much smaller than the Sun can be explained by the new PTA data analyses results \cite{Inomata:2023zup}. Additionally, PBHs that are produced in small abundances are more compatible with PTA observations
 \cite{Depta:2023qst}. 
To summarize, the NANOGrav 15-year results on gravitational waves (GWS)~\citep{NANOGrav:2023gor}, in combination with the \textit{Planck} 2018 data and PBH bounds, can put serious restrictions on the parameters of our model.

%%%%%%%%%%%%%%%%%%%%%%%
%%%%%%%%%%%%%%%%%%%%%%%%%%%%

\section{Conclusion and Discussion}\label{sec5}

In this study, we investigated a mechanism for producing the seed of PBHs in $\mathbb{T}^2$-inflation by examining the coupling between the inflation field and $\mathbb{T}^2$ term. Compared to the standard chaotic inflation, the EMSG term can modify the predictions of the scalar spectral index and the tensor-to-scalar ratio on CMB scales. This modification makes them compatible with the recent BICEP/Keck observational bounds.

 Furthermore, we examined the possibility of enhancing curvature perturbations at specific scales to generate the seed for PBHs, while ensuring that the model remains consistent with CMB observations. As previously discussed, such an enhanced power spectrum leads to PBHs, contributing a large fraction of DM abundance, and simultaneously generating sizable SIGWs. 
 
 The recently published PTA measurements provide evidence of a SGWB. While it aligns with the possibility of a background originating from binary mergers of supermassive black holes, it is intriguing to consider the signal's potential association with the early universe. By tuning the
model parameters, we can observe an enhanced power spectrum in $\mathbb{T}^2$-inflation model at different scales, which enables us to generate primordial black holes (PBHs) with a wide range of masses. Furthermore, it provides us with an opportunity to explain the PTA data via the corresponding SIGWs in our model.
 
 Last but not least, we should comment on the ongoing discussion on quantum loop corrections to the power spectrum in $P(X,\phi) $ theories. It has been noticed that loop corrections put stringent constraints on PBH formation in single field inflation during {\it sharp } slow roll to ultra slow roll transition, namely, enhancement in fluctuation is shifted towards large frequencies, thereby creating PBH with small masses \cite{Firouzjahi:2023aum,Firouzjahi:2023ahg,Kristiano:2022maq,Riotto:2023gpm,Choudhury:2023jlt,Choudhury:2023rks,Choudhury:2023hvf,Choudhury:2023kdb}. It is likely to shift the GW predictions towards the right, to the high frequency regime, which might fall within the LIGO-LISA proposed sensitivities.
 
\section*{Acknowledgments}
We gratefully acknowledge Hassan Firouzjahi for the useful comments and discussions. We thank the partial support from the "Saramadan" Federation of Iran.
MS is partially supported by the Ministry of Education and Science of the Republic of Kazakhstan, Grant No. AP14870191 and CAS President's International Fellowship Initiative(PIFI). A. T. would like to thank University of Rwanda, EAIFR, and ICTP  for their kind hospitalities during the 17th international workshop on the "Dark Side of the Universe" when some parts of the project were in hand.

	\bibliography{PBHT2}

\end{document}